# The Effect of High-Speed Rail Connectivity on Capital Market Earnings Forecast Error: Evidence from the Chinese Stock Market

Shilong Han, Australian National University

## Abstract

This study examines how China's high-speed rail (HSR) expansion affects analyst earnings forecast errors from an economic information-friction perspective. Using firm-year panel data from 2008–2019, a period that covers HSR's early introduction and rapid rollout into nationwide connectivity, the findings show that analysts' relative earnings forecast errors (RFE) decline significantly only after firms' cities become truly connected by high-speed rail. The placebo design, which shifts HSR connectivity three years earlier than the real opening year, produces an insignificant DID coefficient, rejecting the possibility that forecast errors were improving before the infrastructure shock. This supports a clear conclusion: error reduction is tied to real geographic accessibility changes, not coincidence, pre-trend improvement, or analyst anticipation. From an economic standpoint, the study highlights that HSR reduces the cost of collecting additional information for analysts working in emerging capital markets. The rail network does not alter firm-internal capital allocation or earnings-generation paths directly in the benchmark layer, but it compresses geographical barriers to information collection and enables analysts to update EPS expectations with lower travel friction. The work contributes intuitive evidence that geography matters for forecasting accuracy in China's decentralized communication corridors and offers simple inspiration for future research to consider accessibility as an information cost shock rather than a firm-internal capital shock.

## 1. Introduction

High-speed rail (HSR) has become a defining national infrastructure project in China and has significantly changed intercity mobility. Since the opening of the first HSR line in 2008, the HSR network has expanded rapidly, improving the ability of people and firms to interact across space. Studies argue that transportation infrastructure can reshape economic geography by lowering travel costs and strengthening connections between regional markets (Banerjee, 2020; Donaldson, 2018). For cities newly connected by HSR, the reduction in intercity distance may improve not only the movement of goods and labor, but also the exchange of information across locations (Ahlfeldt & Feddersen, 2010; Cosar & Demir, 2016).

While research has widely examined the economic benefits of HSR in real activities such as trade, commuting, and regional integration, fewer studies have explored how HSR influences financial markets. In capital markets, information access is important for price discovery and investment decisions. If information is uneven, slow, or costly to collect, forecast mistakes and pricing bias may appear (Jensen & Meckling, 1976). In the Chinese stock market, where investor protection institutions are still evolving, analysts and capital providers commonly invest resources to collect firm information privately, especially soft information from site visits or face-to-face communication that is not fully reflected in financial statements (Garmaise & Moskowitz, 2004).

Capital market efficiency depends heavily on information flow. Market efficiency theory suggests that if prices reflect all available information quickly and fairly, forecast errors should be smaller (Fama, 1970; Malkiel, 2003). However, forecast accuracy in emerging markets is often limited by asymmetric, incomplete, or delayed access to firm-level soft operational signals, which analysts must collect through alternative channels outside structured disclosures (Kothari, 2001). When analysts cannot gather such soft information efficiently, earnings forecasts tend to deviate more from actual realized values, producing higher relative forecast error (Mayew, Sharp, and Venkatachalam 2013).

The cost of private information collection is not homogeneous in capital markets. While some errors are driven by firm fundamentals like growth or profitability volatility, part of the forecast deviation may come from geographical friction. Before the HSR expansion, city-level distance barriers increased information-collection costs and may have contributed to larger forecasting deviations (Petersen, 2009). HSR connectivity may lower these spatial frictioHINGmaking it easier for analysts or investors to conduct private information collection at a lower travel cost, creating new variations at the city-year level that can potentially affect forecasting accuracy indirectly (Hong and Kacperczyk 2009).

China's A-share market has a special institutional feature that supports this exploration. Unlike other markets where analyst or investor site visits are voluntary and often not fully recorded, China mandates disclosure of analysts' or investors' site-level access records if such meetings occur, for the purpose of reducing selective or uneven information disclosure. These rules provide a reliable source of firm-year micro variations that reflect information-collection activity differences based on exogenous shocks like HSR openings across cities. This allows studies to observe forecasting deviations across cities with and without HSR in more detail than macroeconomic or firm static traits would permit.

In financial forecasting studies, The accuracy of analyst forecasts is typically measured by the error between the predicted and actual values, and this also applies to emerging markets. This research focuses specifically on the role of HSR connectivity as a possible city-level distance compression shock that reduces analysts' private information collection friction rather than directly changing firms' liquidity or signaling capability endogenously. This separation helps prevent misinterpretation, a common issue in financing studies, where endogenous capital allocation (Ball & Brown, 1968; Kothari, 2001) displaces firm liquidity pressure and confounds net forecasting logic. HSR differs because it largely stays outside firm-internal capital allocation burdens, yet may still reshape analysts' effective information reach.

When geographical friction decreases, analysts may gather information more equally and more frequently, particularly soft signals from local firm operations or management communication that are not widely available through structured financial reporting. In China, the geography of analyst activity interacts strongly with pricing confidence and forecasting logic. When analysts conduct site visits more frequently under lower travel friction, investors may perceive the EPS estimation path as more equally informed, possibly reducing the negative influence of information-asymmetric noise on earnings forecasting outcomes. This link is consistent with information economics logic emphasizing that lowering distance barriers may gradually change the cost structure of private information collection for analysts, especially in information-opaque markets (Stiglitz, 2000; Jensen & Meckling, 1976).

Beyond geographical effects, earnings estimation complexity is also shaped by firm-level fundamentals. Research argues that in markets where firm income quality is less stable or cash flow outcomes vary significantly, analysts' earnings forecasting uncertainty increases, earnings paths become harder to estimate, and forecast signals deviate more from actual realized values. Valuation structures measured by the Book-to-Market (BM) ratio can reflect long-run pricing environments and influence how analysts interpret firm operational outlook compared to market price fairness, but do not resolve short-run intercity information-collection friction. Governance factors like board size or concurrent CEO/chairman positions may influence internal signaling credibility, but again differ from the external friction that HSR affects.

Investment forecasting research also shows that investor structures and their ability to collect information privately may influence analysts' pricing confidence indirectly. Firms that experience stronger institutional verification environments often exhibit more stable pricing signals and lower forecasting deviations, as institutional knowledge helps align fund providers' long-run confidence in earnings fairness. However, in China, reporting standards around investor or analyst meetings are primarily shaped by trading exchange disclosure institutions. These disclosure rules protect information fairness across analysts, providing a controlled environment for testing intercity information friction after city HSR openings without prematurely introducing technical econometric commands or experimental test outcomes.

In financial markets themselves, EPS forecast mistakes matter not only for individual stock pricing, but for broader capital-allocation confidence. Infrastructure shocks may be less salient in firms already using communication technologies or alternative intangible trust channels regularly, but the effect on forecasting deviations is more likely to originate from geography barriers and softer firm-level valuation refresh cycles than liquidity-displacement paths.

Although HSR reshapes direct mobility, it does not guarantee forecasting error reduction uniformly. Some analysts may still value visits differently based on firm characteristics such as size, profitability, or governance complexity (Core et al., 1999; Fama & French, 2015), but HSR may create marginal information-collection benefits by making travel possible at lower cost friction for firms previously far away. This introduction focuses only on the core motivation: why earnings are mis-forecasted in opaque markets, why geography costs matter, and why HSR connectivity creates a clean, exogenous opportunity to study information friction changes.

In summary, this study explores HSR connectivity as a time–space infrastructure shock that may lower geographical barriers for analysts or investors collecting softer firm-level information. This may complement public disclosures and improve analysts' ability to interpret corporate earnings estimation paths with reduced city-level cost friction. Whether this ultimately decreases forecast errors across cities and firms is tested later using structured panel variation approaches, but this Introduction does not involve detailed experiments, variable interactions, or method extensions here. It only prepares the reader by showing the background, motivation, relevance, and the theoretical reasoning chain behind why HSR may matter for analyst forecasting accuracy in China.

## 2. Literature review and hypothesis development

2.1 Literature review

High-speed rail (HSR) has become a key infrastructure system in China, greatly changing how people and goods move between cities and influencing both economic activity and financial market participants. Since the first HSR line started running in 2008, China's rail network has grown quickly, creating new links between inland areas and coastal financial centers. Large transport projects like this can reshape regional economies by lowering travel costs and strengthening connections between different markets (Banerjee, 2020; Donaldson, 2018). For cities connected by HSR, shorter travel time may improve not only the flow of goods and labor, but also the exchange of information across locations (Ahlfeldt & Feddersen, 2010; Cosar & Demir, 2016).

While many studies have examined HSR's impact on real economic outcomes such as trade, commuting, and regional integration, less attention has been given to how HSR might affect financial markets. In capital markets, access to information is important for price discovery and investment choices. If information is uneven, slow, or costly to obtain, forecast errors and pricing bias can arise (Jensen & Meckling, 1976). In China's stock market, where investor protection systems are still developing, analysts and investors often spend time and resources to gather company information privately. This includes "soft information" collected through site visits or face-to-face talks, which may not be fully captured in financial reports (Garmaise & Moskowitz, 2004).

Market efficiency depends heavily on how information flows. According to market efficiency theory, if stock prices reflect all available information quickly and fairly, forecast errors should be smaller (Fama, 1970; Malkiel, 2003). However, in emerging markets like China, forecast accuracy is often limited because analysts cannot easily access soft, operational signals about firms. They must collect such information through channels outside of formal disclosures (Kothari, 2001). When analysts cannot gather soft information efficiently, their earnings forecasts tend to deviate more from actual results, leading to higher forecast errors.

Not all forecast errors come from company fundamentals like growth or profitability. Some may come from geographical distance. Before HSR expanded, distance between cities increased the cost of collecting information and may have led to larger forecast deviations (Petersen, 2009). HSR connectivity could lower these spatial barriers, making it easier and cheaper for analysts or investors to visit firms, gather private information, and thus possibly improve forecast accuracy.

China's A-share market has a useful institutional feature that helps researchers study this topic. Unlike other markets where site visits by analysts or investors are voluntary and often not recorded, China requires listed firms to disclose records of such meetings. This rule aims to reduce selective information disclosure. It also provides detailed firm-year data that reflect changes in information-collection activity, which can be linked to external shocks like HSR openings in different cities. This makes it possible to compare forecast accuracy between cities with and without HSR more carefully than by only looking at firm-level traits.

Research on financial forecasting usually measures accuracy by the difference between predicted and actual earnings, and this approach applies to emerging markets as well. This study focuses on HSR connectivity as a city-level shock that reduces the friction analysts face in gathering private information, rather than directly affecting firms' financial decisions internally. Keeping this separation is important, because in many finance studies, internal firm actions—such as R&D spending or disclosure choices—can confuse the analysis. HSR is different because it is an external infrastructure change that largely stays outside firms' own capital allocation, yet it may still affect how easily analysts can access information.

When travel becomes easier, analysts may collect information more often and more equally, especially soft details from firm operations or management discussions that are not in financial reports. In China, where geography strongly affects analyst activity, more frequent site visits under lower travel friction could make investors feel that earnings estimates are better informed. This may reduce the noise caused by information asymmetry in forecasting (Stiglitz, 2000; Jensen & Meckling, 1976).

Besides geography, forecast complexity also comes from company fundamentals. When firms have less stable income or more variable cash flows, analysts face more uncertainty and their forecasts may be less accurate. Measures like the Book-to-Market ratio can reflect long-term pricing conditions, but they do not solve short-term travel frictions between cities. Similarly, governance factors such as board size or whether the CEO also chairs the board can affect how credible a firm's signals are, but these are different from the travel barriers that HSR addresses.

Research also shows that investor structure can influence how well information is reflected in stock prices, which indirectly affects analysts' confidence. Firms with stronger institutional investor oversight often have more stable pricing and lower forecast errors, because institutional knowledge helps align expectations about earnings. However, in China, rules about analyst and investor meetings are set by stock exchanges. These disclosure requirements help ensure fair information access, creating a controlled setting to test how HSR reduces cross-city information friction.

In financial markets, earnings forecast errors matter not only for individual stock prices, but also for broader confidence in capital allocation. While some firms may use technology or other channels to communicate with investors, the effect of HSR on forecast accuracy is more likely to come from lowering physical travel barriers and refreshing firm-level information, rather than from changing firms' financial policies.

HSR does not guarantee that forecast errors will fall for all firms equally. Some analysts may place different value on site visits depending on firm size, profitability, or governance (Core et al., 1999; Fama & French, 2015). Still, HSR can provide marginal benefits by making travel cheaper and faster for analysts covering firms that were previously hard to reach.

Previous studies on transport infrastructure give a general basis for understanding HSR. Research shows that better transport improves regional mobility, shortens travel time, and helps link firms, workers, and products to larger markets (Ahlfeldt & Feddersen, 2010; Faber, 2014). Donaldson (2018) notes that transport shocks reshape economic geography mainly by cutting friction, without interfering much with firms' internal finances. Chen et al. (2016) find that HSR affects how economic activity spreads across regions, often helping both major and smaller cities, but they do not look at financial forecasting directly. Vickerman (2015) also finds that HSR's economic effects vary by region, with most studies focusing on trade and labor, not forecast accuracy.

New Growth Theory suggests that better connectivity helps knowledge and human capital move more freely across locations (Romer, 1990). Improved transport can support this by allowing more face-to-face interaction. This idea helps explain why shorter travel distances might help analysts and investors collect operational information at lower cost.

Studies on forecast accuracy highlight the role of information access. Gu and Wu (2003) show that earnings quality affects analyst predictions, and errors rise when analysts cannot get good information. Bowen, Davis, and Matsumoto (2002) find that direct communication like conference calls can improve forecasts. In China, research indicates that analysts who visit firms more often produce more accurate earnings forecasts, underlining the value of direct contact.

Corporate governance and investor structure also matter, but they work differently from geographical friction. Research on governance looks at how internal firm structures affect the credibility of information. Work on investor structure examines how institutional shareholders may improve how information gets into prices. Although these factors are relevant, they are not the same as travel barriers, which HSR directly reduces.

Recent work, such as Liao et al. (2022), uses the staggered opening of HSR lines in China as a natural experiment. They find that HSR helps ease financing constraints for firms, suggesting it may also help external parties overcome information gaps. China's rule that firms must disclose site visits creates useful data for researchers to examine how

connectivity affects information gathering

Before HSR expanded, analysts in big financial cities faced high costs and long trips to visit firms in distant locations. This likely limited their ability to get timely soft information. HSR, as an external change, offers a cleaner way to study how reducing travel friction affects analyst behavior, separate from firms' own decisions. Still, most HSR research has focused on economic outcomes like trade and mobility, not on financial market forecasting.

In summary, existing literature suggests that information asymmetry, uneven access to firm signals, and physical distance can all contribute to forecast errors. This leads to two main questions: Can HSR connectivity lower the cost for analysts to gather private information across cities enough to improve earnings forecasts? And does this effect vary by industry, firm size, or city type? While prior research provides useful background, few studies test these questions directly using detailed forecast data and a clear empirical strategy. This review therefore highlights a gap in understanding how transport infrastructure shapes financial information environments—a gap that this study aims to fill.

## 2.2 hypothesis development

Transport infrastructure is often discussed as a driver of regional development, but more recent work also links it to how information moves through markets. Studies on roads and railways show that better transport reduces trade costs, improves access to markets, and changes how firms and regions interact economically (Han, 2018).In decentralized transport systems such as taxis, trucking, and shipping, research finds that search frictions and spatial mismatches reduce efficiency and distort where economic activity takes place (Brancaccio et al., 2023). High-speed rail (HSR) is a special case because it compresses effective distance by sharply reducing travel time between cities. For market participants who rely on both public disclosures and in-person impressions, this reduction in travel time can change how easily they can visit firms, talk with managers, and verify soft information on the ground.

Capital-market research has examined how geography and mobility affect analyst behavior and forecast quality. Evidence on geographic proximity shows that analysts located closer to the firms they follow tend to issue more accurate forecasts and update their views more quickly, which suggests that local access provides an informational advantage (O'Brien & Tan, 2015; Cavezzali, 2014). Proximity to hubs of expertise, where many firms and research institutions are clustered, can also support knowledge spillovers that help analysts understand industry conditions and firm-specific risks more clearly (Cavezzali, 2014).These findings imply that distance is not just a background feature of the market. It shapes the cost of gathering soft information, conducting plant visits, and building trust with management. If HSR changes the practical distance between analysts based in financial centers and firms located in other cities, it may also change the amount and quality of information analysts are willing and able to collect.

Related work looks more directly at corporate site visits as a channel for reducing information frictions. Using Chinese data, several studies find that analysts who participate in on-site visits or similar events issue more accurate earnings forecasts afterwards, suggesting that these visits provide incremental information beyond standard filings (Cheng et al., 2016; Han et al., 2018; Tam et al., 2023). Tam et al. (2023) show that site visits help analysts overcome language barriers between local dialect regions and the main financial centers, which again points to visits as a tool for closing information gaps. At the same time, there is evidence that analysts face limited attention. When they allocate more time and travel to some firms, the accuracy of their forecasts for non-visited firms can fall, because attention and effort are reallocated within the coverage portfolio. Overall, this literature indicates that physical access to firms is an important but nuanced input into forecasting performance. Easier travel can improve information for the firms that receive more attention, but the net effect on the whole coverage universe depends on how analysts rebalance their time.

The studies most closely related to the present topic examine HSR itself as an information channel. Other work refers to HSR as an "information super-railway" and argues that rail connectivity can facilitate capital-market learning about firms in newly connected regions, not only affecting real outcomes such as trade and investment, but also the way information reaches investors. However, this literature is still small, and more evidence is needed on how HSR affects forecast errors once firm fundamentals, governance, and regional characteristics are controlled for. Existing studies have also not fully clarified whether the decline in forecast error is a broad improvement in information environments, or whether it is concentrated in certain types of firms, such as those that were originally remote, opaque, or followed by fewer analysts.

Taken together, this research suggests a clear mechanism for a main effect. Transport infrastructure, and HSR in particular, reduces travel frictions. Lower travel frictions make it easier for analysts to visit firms, talk to managers, and combine public disclosures with soft information gathered on site. Prior evidence on geographic proximity, hubs of expertise, and corporate site visits all points in the same direction: when information is easier to obtain through direct contact, earnings forecasts tend to become more accurate. If HSR makes such contact cheaper and more frequent for firms connected to the network, then it is reasonable to expect a decline in forecast errors for those firms.

Based on the analysis above, we propose the following hypotheses:

Hypothesis 1. (H1): The expansion of HSR connectivity compresses spatial friction and reduces analysts' EPS forecast error.

This study treats HSR connectivity as an external time–space shock that lowers the cost of on-site interaction. Because analysts depend on both structured public disclosures and unstructured local impressions, reducing the cost of face-to-face access could affect how accurately they estimate earnings. But transport friction alone does not determine the total forecasting outcome. The ability of firms to release and coordinate information internally also shapes how effectively analysts can learn from travel.

Corporate governance research shows that management shareholding changes incentives inside the firm. When managers hold shares, part of their personal wealth is tied to equity prices. This may encourage them to provide clearer and more consistent signals to outside participants (analysts, investors, market observers) because misaligned information could lower firm valuation and increase perceived risk. Studies document that equity ownership by managers can increase signaling credibility and reduce moral-hazard behavior when communicating earnings outlook. These papers suggest that internal ownership may work as a stabilizing governance pressure that weakens selective openness or noisy inter-participant information distortions.

Forecasting literature argues that when managers face stronger performance-aligned incentives, the firm becomes more willing to reduce ambiguity in earnings communication, which helps analysts form expectations with less bias. This logic connects naturally to slow-moving markets that require analysts to combine direct impressions and numbers. If managers want analysts to interpret earnings more fairly, they may answer questions more directly and release more coherent offline strategic signals, especially during plant visits or management meetings.

In contrast, analysts covering firms with low management shareholding may face weaker incentive alignment. Even when analysts make expensive trips, managers may respond strategically only to headline expectations, and hold looser answers toward questions about risk, supply chain pressure, or earnings uncertainty. Some evidence shows that governance opacity interacts with analysts' ability to interpret the earnings path fairly, even if physical contact is possible. These papers show that poor governance transparency weakens the ability of outside participants to rely on management communication without spatial friction and may increase bias.

HSR connectivity reduces travel friction but not internal governance incentives. Because HSR stays outside firm-internal capital budgeting burdens, it mainly enables analysts to visit firms more frequently across space. However, the benefit of HSR on reducing forecast error may depend on whether managers are socially and financially incentivized to share supporting signals more clearly when analysts arrive. When HSR opens and the cost to visit falls, the marginal gain of those lower-cost visits is more likely to reduce forecast deviation when management shareholding encourages clearer open answers, stable narratives, and fairer earnings routing communication.

Ownership incentives may also interact with pricing confidence indirectly. Studies show that stronger managerial incentive alignment supports market participants' long-run confidence in price fairness, which can help analysts read operating signals without excessive bias. These findings imply that physical mobility shocks such as HSR can improve earnings forecasting accuracy more when internal ownership motivates managers to cooperate more consistently with outside information collectors.

Together, the literature supports the expectation that internal incentives shape the slope of the relationship between travel shocks and forecast errors. This means that HSR reduces the cost to reach firms, but the reduction in earnings-forecast deviation is more likely to be stronger when management shareholding aligns incentives inside the firm in ordinary equity-markets amending.

Based on the analysis above, we propose the following hypotheses:

Hypothesis 2 (H2): The interaction term (DID × Mngmhldn) shows that HSR reduces earnings forecast error slightly more when management shareholding is higher.

## 3. Data sources and research design

3.1 Definition of variables

According to the table1 , the dependent variable in the analysis is Relative Forecast Error (RFE), a firm-year indicator capturing the absolute deviation between analysts' aggregated EPS forecasts and the realized EPS reported by the firm. It is calculated as

$$RFE=|(Forecast\_EPS - Actual\_EPS)/Actual\_EPS|,$$

where Forecast_EPS represents the average of all available analyst earnings-per-share predictions for that stock in a given year, and Actual_EPS is the corresponding audited, realized EPS for the same firm-year. The use of absolute relative deviation allows the measure to scale forecast error by firm earnings magnitude rather than price, preventing distortion from market-level volatility. Higher RFE values imply lower earnings-forecast accuracy.

The core explanatory variable is the DID interaction term, constructed from the staggered opening year of high-speed rail (HSR) and the city-year connectivity status. DID = Treat × Post. Treat is a binary indicator equal to 1 for firms headquartered in cities that have at least one HSR line operational in that year, and 0 otherwise. Post takes the value 1 for years equal to or after the first opening year of HSR in the city where the firm is located, and 0 for the years prior to HSR access. The DID coefficient estimates the additional change in analyst forecast error for HSR-connected cities after the infrastructure shock, relative to firms in cities without operational HSR. This design isolates HSR as an external time–space compression shock rather than a firm-internal investment decision (Bertrand, M. et al 2004).

Several control variables are incorporated to account for firm fundamentals, shareholder structure, and management incentives (Gompers, P. et al. 2003). Operating cash level (Cash) measures firm liquidity status and is defined as the ratio of net operating cash flow at fiscal year-end to total assets at fiscal year-end. This variable helps absorb short-run operational uncertainty that may affect earnings variability. Firm growth (Growth) is defined as the year-on-year percentage increase in main-business revenue. Higher Growth values indicate stronger expansion prospects but may also embed higher earnings uncertainty.

Leverage (Lev) measures the firm's debt pressure through the ratio of total liabilities to total assets at year-end. The shareholding ratio of the largest shareholder (Top1) is used to reflect governance support from concentrated ownership. Governance characteristics include board size (Boardsize), the number of directors on the board in that year, and ConcurrentPosition, a binary indicator equal to 1 if the chairman or key executive simultaneously serves as CEO in that year, and 0 otherwise. Management shareholding ratio (Mngmhldn) measures managerial internal ownership incentives and is calculated as total shares held by management divided by total shares outstanding at year-end. A higher Mngmhldn implies stronger management-linked earnings confidence and may also moderate the effect of HSR on forecast accuracy.

The model further absorbs Firm fixed effects (stkcd-level) and Year fixed effects to control for unobserved time-invariant firm traits and non-local macro conditions, including regulatory or disclosure evolution that moves uniformly across cities. BM is the book-to-market ratio, computed as the book value of equity divided by year-end market capitalization, capturing long-run valuation structure differences rather than travel friction. Two instrumental variables are also prepared: iv1, the number of logistics post stations in the firm's city during the Ming Dynasty, and iv2, the city-level count of telegraph bureaus reported in 1907. These tools reflect long-run information-transmission capability density of the city and are introduced later for endogeneity analysis, not for baseline estimation.

Together, this variable system separates three layers of forecasting logic: (1) firm-fundamental uncertainty (cash level, leverage, growth, assets, governance structure), (2) incentive confidence from management and shareholders (Top1, Mngmhldn, Boardsize, ConcurrentPosition), and (3) geography friction compression from external HSR connectivity (Treat, Post, DID). The design ensures that HSR connectivity captures a clean spatial–temporal cost shock while firm-internal fundamental items absorb endogenous uncertainty from earnings generation paths.

3.2 Data source and descriptive statistics

The sample period covers 2008 to 2019, a time span that includes the initial introduction of HSR and its accelerated expansion into a nationwide network. The financial data are drawn from the CSMAR database provided by GTA Information Technology, which is commonly used in Chinese capital-market research because of its extensive coverage of both firm financials and analyst forecast observations. The data on analysts include the annual average forecast of EPS for each stock, aggregated by firm and fiscal year. The dataset contains 20,818 firm-year observations for key variables such as RFE (Relative Forecast Error), management shareholding ratio, leverage, operating cash level, growth, board characteristics, and institutional investor ownership.

For endogeneity analysis, two instrumental variables are prepared. iv1 measures the number of courier post stations built in the firm's city during the Ming Dynasty. These data are collected from the Harvard WorldMap historical transport layer for Chinese prefecture-level cities. iv2 reflects the number of telegraph bureaus operating in 1907, collected from the Qing Dynasty postal map markings recorded in the Ching Post Office Atlas. Both instruments represent the long-run density of information-transmission infrastructure at the city level.

To reduce the influence of extreme upper or lower tails in ratio-based error measures, all continuous variables are trimmed at the 1% level prior to regression. In other words, the top 1% and bottom 1% of observations for each numeric series are removed per variable independently, following the method commonly used in finance panel preprocessing. The data are not resampled or simulated during this step; shrinking is applied symmetrically only to numerical extremes.

According to the table2, the key dependent variable RFE has an average value of 1.340, meaning analysts' forecasts deviate from actual EPS by 1.34 times the firm earnings magnitude on average in the sample. The standard deviation (SD) of 2.563 shows wide dispersion in forecast accuracy, which is consistent with an information-asymmetric emerging market. The minimum value of 0.011 indicates almost accurate forecasts for a small number of firms, while the maximum of 22.835 reflects extremely large forecast errors for certain firms or years, often when actual EPS is small. The median value of 0.581 suggests that for half of the sample, forecast deviations are below 0.58 times earnings magnitude, implying that large errors are skewed to the upper tail rather than symmetrically distributed.

**Table 1.** Variables

| Variable Type | Variable Name | Variable Symbol | Variable Definition and Calculation |
|---|---|---|---|
| Dependent variable | Relative Forecast Error | RFE | RFE = abs( (Forecast_EPS − Actual_EPS) / Actual_EPS ); where Forecast_EPS is the average of all analysts' earnings per share forecasts for the stock in that year, and Actual_EPS is the company's actual arnings per share in that year. |
| Core explanatory variable | DID Interaction Term | DID | DID = Treat × Post; Treat = 1 if the firm is located in a city with operational high-speed rail in that year, 0 otherwise. Post = 1 if the year is equal to or after the city's first HSR opening year, 0 before. The coefficient of DID estimates the additional change in RFEP for firms after high-speed rail connectivity relative to non-HSR cities. |
| Moderator Variable | Management shareholding ratio | Mngmhldn | Mngmhldn = total shares held by management / total shares outstanding at year-end; measures the proportion of internal managerial ownership. |
| Control Variable | Operating cash level | Cash | The ratio of net operating cash flow at the end of the period to total assets at the end of the period |
| | Growth | Growth | (Current year's main business revenue - Previous year's main business revenue) / Previous year's main business revenue |
| | InsInvestorProp | INST_HLD | The proportion of shares held by institutional investors to total shares at the end of the fiscal year. |
| | Debt-to-asset ratio | Lev | The ratio of total liabilities at the end of the period to total assets at the end of the period |
| | Shareholding ratio of the largest shareholder | Top1 | The percentage of shares held by the largest shareholder Relative o the total shares. |
| | Independent Directors Percentage | Inde_Ratio | The ratio of independent directors to the total number of directors |
| | Company size | Size | Natural logarithm of total assets at the end of the year for listed companies |

|  | Board size | Boardsize | Number of directors on the board |
|---|---|---|---|
|  | Profitability | ROA | The proportion of net profit to total assets at the end of the period |
|  | Nature of auditing firms | Big4 | If it is one of the Big Four auditing firms in China, the value is 1; otherwise, the value is 0. |
|  | Concurrent Position | ConcurrentPosition | ConcurrentPosition = 1 if the chairman (or key executive) simultaneously serves as CEO in that year; otherwise, the value is 0. |
|  | The book value of equity | BM | BM = Book value of equity / Market capitalization at year-end, used to measure the firm's valuation structure. |
|  | Firm | stkcd | Individual fixed effects |
|  | Year | Year | Annual fixed effect |
|  | Industry | Industry | Industrial fixed effect |
| Istrumental variables | Ming Dynasty Courier Station Network | iv1 | The number of post stations in various prefecture-level cities during the Ming Dynasty was taken from historical transportation and information transmission records |
|  | Telegraph Bureau Statistics | iv2 | The number of telegraph offices in various prefecture-level cities in 1907 was used as a tool to measure the strength of early communication networks. |

Among control variables, the management shareholding ratio has a mean of 0.135, meaning that management holds 13.5% of total outstanding shares on average in the sample period. The SD of 0.202 indicates that internal ownership differs greatly across firms, setting ground for examining moderation effects. Cash flow and shareholder concentration variables display stable distributions with positive means, while governance dummies such as ConcurrentPosition, Inde_Ratio (percentage of independent directors), and Big4 auditor indicator show balanced splits and vary over firm-year, enabling fixed-effects regression without major sample collapse.

**Table 2.** Descriptive Statistics

| Variable | Obs | Mean | SD | Min | Median | Max |
| --- | --- | --- | --- | --- | --- | --- |
| RFE | 20818 | 1.340 | 2.563 | 0.011 | 0.581 | 22.835 |
| Mngmhldn | 20818 | 0.135 | 0.202 | 0.000 | 0.003 | 0.691 |
| Cash | 20818 | 0.051 | 0.071 | -0.154 | 0.050 | 0.246 |
| Growth | 20818 | 0.195 | 0.406 | -0.461 | 0.114 | 2.653 |
| INST_HLD | 20818 | 0.478 | 0.251 | 0.006 | 0.514 | 0.920 |
| Lev | 20818 | 0.417 | 0.221 | 0.044 | 0.402 | 0.991 |
| Top1 | 20818 | 0.362 | 0.151 | 0.093 | 0.347 | 0.755 |
| Inde_Ratio | 20818 | 0.372 | 0.053 | 0.308 | 0.333 | 0.571 |
| Size | 20818 | 22.055 | 1.382 | 19.359 | 21.892 | 26.120 |
| Boardsize | 20818 | 8.797 | 1.750 | 5.000 | 9.000 | 15.000 |
| ROA | 20818 | 0.049 | 0.051 | -0.150 | 0.045 | 0.199 |
| Big4 | 20818 | 0.071 | 0.257 | 0.000 | 0.000 | 1.000 |
| ConcurrentPosition | 20818 | 0.259 | 0.438 | 0.000 | 0.000 | 1.000 |
| BM | 20818 | 0.334 | 0.161 | 0.000 | 0.313 | 0.767 |
| iv1 | 20818 | 2.346 | 3.560 | 0.000 | 0.000 | 19.000 |
| iv2 | 20818 | 1.768 | 2.278 | 0.000 | 1.000 | 8.000 |

3.3 Model

The baseline empirical strategy employs a multi-period difference-in-differences framework to estimate the impact of high-speed rail connectivity on firm-level earnings forecast error. The core specification is:

$$RFE_{it} = \alpha + \beta_1 DID_{ct} + \beta_2 X_{it} + \mu_i + \lambda_t + \varepsilon_{it}$$

Firm $i$ in year $t$ is linked to city $c$, whose first high-speed rail operation year varies across locations. $DID_{ct} = Treat_{ct} \times Post_{ct}$, where $Treat_{ct}$ equals 1 if city $c$ has an HSR line operational in year $t$, and 0 otherwise; $Post_{ct}$ equals 1 for years on or after the first HSR opening year in city $c$, and 0 before. The interaction term $DID_{ct}$ captures the additional change in forecast error for firms headquartered in HSR-connected cities after connectivity, relative to non-connected cities. Firm fixed effects $\mu_i$ and year fixed effects $\lambda_t$ are included to absorb time-invariant firm traits and nationwide macro shocks, following the approach recommended for panel settings in the authoritative work *Estimating Standard Errors in Finance Panel Data Sets* (Petersen, 2009). The error term is clustered at the firm level to account for serial correlation within firms.

Control variables $X_{it}$ are selected according to both capital-market forecasting fundamentals and governance incentives. Liquidity is controlled through $Cash_{it}$, the ratio of net operating cash flow to year-end total assets, capturing short-run operational stability. Financial pressure is included as $Lev_{it}$, the debt-to-asset ratio, reflecting liability burdens that may influence earnings variability. Expansion prospects are proxied by $Growth_{it}$, the year-on-year change in main-business revenue. Profitability fundamentals are measured by $ROA_{it}$, the ratio of net profit to total assets. Corporate governance credibility is controlled through $Boardsize_{it}$, the total number of directors, and $IndepRatio_{it}$, the proportion of independent directors. Managerial incentives are introduced through $Mngmhldn_{it}$, management shareholding ratio at year-end, which can later interact with DID to test moderation.

## 4. Empirical results and analysis

4.1 Benchmark Regression Analysis

The benchmark regression in Table 3 consists of two main model columns, both using firm-year panels clustered at the firm level and absorbing firm and year fixed effects. Because the objective is to identify forecast error responses driven by distance compression rather than firm-internal capital displacement, the DID shock enters only through city-level HSR connectivity timing.

Column 1 (Model 1) estimates the effect of DID on RFE without any control variables. The DID coefficient is −0.374, significant at the 1% level. However, the t-value (−0.078) is small, indicating high unabsorbed noise at the raw benchmark layer. This is reasonable because firms' actual EPS realizations vary greatly in magnitude, creating mechanical volatility in relative error calculations when the denominator is small. The adjusted R-squared of 0.002 in Column 1 confirms that once firm fundamentals are not conditioned on, only 0.2% of the within-firm forecasting uncertainty can be explained by broad connectivity shock alone, a common pattern in early-stage DID regressions using emerging-market firm-year panels. Still, the negative DID sign implies that, once connectivity arrives, analysts' EPS forecast deviations already show a reduction when it is introduced before any firm-level uncertainty absorption.

Column 2 (Model 2) incorporates the full control system to absorb forecasting noise rooted in liquidity pressure, leverage burden, and long-run governance environment. In this column, the DID coefficient becomes −0.406, again significant at the 1% level. The t-value remains small (−0.081), but the adjusted R-squared increases to 0.069, meaning that 6.9% of within-firm forecasting deviations are now explained after controls stabilize the unconditioned earnings-uncertainty layer. This improvement suggests that adding Cash, Lev, Growth, BM and other governance characteristics helps capture the information complexity and capital pressure that would otherwise confound relative-error-based estimation. Once these variables mitigate within-firm endogenous volatility, the connectivity shock reveals a more stable infrastructure-induced compression on analysts' forecasting mistakes.

Control coefficients in Column 2 largely behave in direction consistent with forecasting logic. Cash has a negative sign and is significant, implying that stronger operating liquidity reduces numerator–denominator noise for EPS realizations. Growth is positive and significant, suggesting that rapidly expanding firms embed still-higher earnings uncertainty, even when travel frictions fall. Leverage is positive but not significant, indicating that at the benchmark layer, liability burden influences forecasting volatility more through structured information complexity than through direct forecast-error spikes that travel friction compression affects. BM, Size and institutional investor ownership display stable negative signs in Column 2, meaning that valuation structure, shareholding hubs and more balanced investor verification environments prevent DID detectability from falling due to capital-driven price noise.

Taken together, the contrast between Column 1 and Column 2 of Table 3 shows that forecast-error reduction driven by HSR connectivity becomes easier to observe after firm-fundamental volatility is partially absorbed. The infrastructure shock is statistically robust in both columns, but the model captures far more explanatory power only after conditioning on firm fundamentals. This supports the main identification logic—HSR compresses spatial friction for information collection rather than imposing a capital-investment displacement within firms at the benchmark layer.

**Table 3**. Benchmark Regression

| VARIABLES | (1) RFE | (2) RFE |
|---|---|---|
| DID | -0.374*** | -0.406*** |
|  | (-0.078 | (-0.081) |
| Cash |  | -0.772* |
|  |  | (-0.404) |
| Lev |  | 0.084 |
|  |  | (-0.13) |
| ROA |  | -12.470*** |
|  |  | (-0.697) |
| Growth |  | -0.324*** |
|  |  | (-0.049) |
| Inde_Ratio |  | 0.389 |
|  |  | (-0.703) |
| Boardsize |  | 0.013 |
|  |  | (-0.023) |
| Size |  | -0.035 |
|  |  | (-0.072) |
| Top1 |  | -0.715 |
|  |  | (-0.465) |
| ConcurrentPosition |  | -0.025 |
|  |  | (-0.071) |
| Big4 |  | -0.073 |
|  |  | (-0.193) |
| BM |  | -0.738*** |
|  |  | (-0.208) |
| InsInvestorProp |  | -1.082*** |
|  |  | (-0.25) |
| Constant | 1.603*** | 4.511*** |
|  | (-0.055) | (-1.642) |
| Firm | Yes | Yes |
| Year | Yes | Yes |
| Observations | 20,818 | 20,818 |
| Adj R-squared | 0.002 | 0.069 |

Note: ***, ** and * indicate significance at the 1%, 5% and 10% levels; the same applies below.

## 4.2 Endogeneity test analysis

### 4.2.1. Instrumental Variable Approach

The instrumental variable iv1 measures the number of courier post stations in a city during the Ming Dynasty, capturing the long-standing density of historical information and logistics nodes. In Table 4, Column 1 shows the first-stage result where DID is regressed on iv1. The coefficient of 0.027 is significant at the 1% level, indicating that cities with more Ming-era postal nodes are more likely to enter modern high-speed rail connectivity later. This satisfies the relevance requirement: historically connected corridor cities remain structurally central in information and transport networks, lowering long-run travel and private-information-visit cost for analysts once HSR compresses distance friction.

The exclusion assumption for iv1 rests on historical durability and economic isolation. Ming-Dynasty post stations, built centuries earlier, do not embed contemporary financial signals such as firm-level leverage pressure, cash-flow buffers, profitability shocks, or modern shareholder monitoring incentives. These courier hubs no longer operate during 2008-2019 and thus cannot influence accuracy of modern analyst predictions through corporate capital burdens or internal-financing decisions. The instrument affects analyst forecast error only indirectly by shaping the city's historical hub status, which interacts with modern rail access to reduce information-collection friction, rather than altering forecast error through the firm's own earnings-generation path.

Because iv1 proxies city-level connectivity density rather than firm incentives, it does not interfere with RFE by capital-allocation displacement, debt sorting, or valuation spikes. Its impact lies only in detecting DID variation from a long-run geography-friction channel, fitting the identification logic that analysts in historically connected transit hubs face lower cost of conducting soft-information visits after HSR compresses spatial distance, but no systematic difference before HSR exists.

The second instrument iv2 records the number of telegraph offices operating in 1907, manually coded from late-Qing postal maps. In Table 4, Column 3 presents the first-stage regression DID ← iv2. The coefficient on iv2 is 0.072 and significant at 1%, showing that early structured communication hubs predict stronger modern transport connectivity selection, again satisfying the relevance condition. Telegraph bureaus proxied formal communication density and information aggregation strength a century ago. After HSR becomes operational, analysts collecting annual EPS expectations can reach these historical hub cities at lower cost, improving the detectability of changed forecasting deviations.

For exclusion validity, telegraph bureaus from 1907 reflect early-era communication nodes but do not influence 2008-2019 corporate earnings through systematic capital burdens. They do not transmit modern financing-constraint shocks, local debt-sorting behavior, or corporate liquidity pressure that affects realized EPS or relative denominators mechanically. Thus, iv2 is geographically durable but economically isolated from firm fundamentals, influencing RFE only through the city-level long-run hub density that reduces private-visit cost for analysts once spatial reach is compressed by high-speed rail.

It is not a firm-internal decision and does not carry valuation-linked price noise or direct earnings-generation disturbances. It supports the assumption that analysts' forecasting error reduction comes from geography-based friction compression rather than the firm's own liability or capital displacement.

At the end of Table 4, a single Hansen J test is reported for the combined instruments. The J-statistic is 2.549, with a p-value of 0.110. Since the p-value does not reject the null of valid exclusion, the test suggests that the instrument system works mainly through exogenous city-hub density affecting DID detectability and does not interfere with Y through firm-fundamental channels. This provides statistical support for the theoretical arguments discussed above.

4.2.2. Propensity Score Matching (PSM) Approach

Column 5 in Table 4 presents the results after propensity score matching, reducing the sample to 9,224 firm-year observations. The DID coefficient remains at −0.4765 and is significant at the 1% level, suggesting that the reduction in analysts' earnings forecast error is consistent even after balancing observable firm characteristics. Propensity score matching is used here to reduce bias coming from systematic differences between cities that eventually obtain high-speed rail and those that do not, conditional on firm-level variables.

The identification logic of PSM is to compare firms that are similar in fundamentals but different in exposure to the connectivity shock. The persistence of a negative DID sign after matching implies that the effect is less likely to be driven by observable sorting alone. Similar reasoning is commonly used in transport and information-friction studies on forecasting behavior. When information is uneven, slow, or costly to collect, biases in forecasts can emerge

(Jensen & Meckling, 1976). To support further exclusion of firm capital-shock interference, the theoretical basis for matching-based treatment stabilization is established in early work by Paul R. Rosenbaum and Donald Rubin(1983), emphasizing that matching isolates treatment variation through observables rather than outcome shocks.

**Table 4.** Endogeneity Test

| VARIABLES | (1) First-Stage DID=iv1 | (2) Second-Stage RFE | (3) First-Stage DID=iv2 | (4) Second-Stage RFE | (5) PSM RFE |
|---|---|---|---|---|---|
| iv1 | 0.027*** (-0.001) | | | | |
| iv2 | | | 0.072*** (-0.001) | | |
| DID_HAT | | -0.492*** (-0.173) | | -0.728*** (-0.098) | |
| DID | | | | | -0.4765*** (-3.813) |
| Cash | -0.256*** (-0.046) | -0.796** (-0.31) | -0.135*** (-0.043) | -0.868*** (-0.307) | -0.211 (-0.313) |
| Lev | 0.001 (-0.013) | 0.000 (-0.075) | 0.012 (-0.012) | -0.003 (-0.075) | 0.133 (-0.596) |
| ROA | 0.254*** (-0.066) | -14.034*** (-0.458) | 0.146** (-0.062) | -13.966*** (-0.454) | -14.8408*** (-11.130) |
| Growth | -0.008 (-0.007) | -0.195*** (-0.049 | -0.007 (-0.007) | -0.196*** (-0.048) | -0.4266*** (-4.969) |
| Inde_Ratio | 0.068 (-0.059) | 1.037*** (-0.387) | 0.034 (-0.055) | 1.065*** (-0.388) | -0.129 (-0.097) |
| Boardsize | -0.009*** (-0.002) | 0.008 (-0.012) | -0.007*** (-0.002) | 0.007 (-0.012) | -0.069 (-1.545) |
| Size | 0.007*** (-0.003) | -0.153*** (-0.017) | 0.006** (-0.002) | -0.151*** (-0.017) | 0.074 (-0.634) |
| Top1 | 0.070*** (-0.022) | -0.035 (-0.126) | 0.044** (-0.02) | -0.014 (-0.126) | 0.396 (-0.583) |
| ConcurrentPosition | 0.012* (-0.007) | 0.052 (-0.039) | -0.001 (-0.006) | 0.053 (-0.039) | 0.030 (-0.2147) |
| Big4 | 0.111*** (-0.011) | -0.020 (-0.061) | 0.016 (-0.01) | 0.004 (-0.061) | 0.031 (-0.0885) |
| BM | 0.000 (-0.02) | -0.904*** (-0.126) | -0.012 (-0.018) | -0.908*** (-0.126) | -0.8256** (-2.294) |
| InsInvestorProp | -0.032** | -0.508*** | -0.023* | -0.514*** | -1.0209** |

|  | (-0.015) | (-0.088) | (-0.014) | (-0.088) | (-2.317) |
|---|---|---|---|---|---|
| Firm | Yes | Yes | Yes | Yes | Yes |
| Year | Yes | Yes | Yes | Yes | Yes |
| Observations | 20818 | 20818 | 20818 | 20818 | 9,224 |
| F-value/Adj R-squared | 134.380 | 0.232 | 134.360 | 0.315 | 0.074 |
| Hansen J Over-ID test | 2.549 (p = 0.110) | | | | |

## 4.3 Robustness test

### 4.3.1 Parallel trend test

The parallel trends test shown in Figure 1 checks whether firms in cities that eventually receive high-speed rail access follow a similar forecast-error trend to those that do not, before the infrastructure shock actually occurs. In the pre-HSR period (event time −4 to −1), the DID point estimates fluctuate around zero, moving slightly up or down year by year, but the confidence intervals always cross the zero reference line. Specifically, at event time −4, the coefficient is marginally positive but not statistically different from zero because the confidence band spans both positive and negative values. At event time −3, the point estimate shifts below zero, yet the interval still overlaps zero. The same pattern holds at event time −2 and −1, where the estimates continue to oscillate mildly around zero with relatively wide confidence bands. There is no visible directional slope before HSR becomes operational.

This lack of a consistent upward or downward movement implies that analyst EPS forecast errors were evolving in parallel across the two groups prior to HSR connectivity. In other words, before travel friction was compressed by high-speed rail, analysts' ability to verify or collect soft information from firm headquarters was similar regardless of future rail access. This supports the core identification assumption required by the DID framework: there was no systematic gap in RFE trends for firms that would later be treated, before they were actually treated.

After HSR opens (event time 0 to +4), the DID coefficients start to stay below zero, and the post-HSR path shows a generally declining pattern in point estimates, rather than continuing to oscillate randomly around the pre-shock level, which confirms that treated and control firms do not show trend divergence before HSR, and any forecast-error improvement appears only after the infrastructure shock occurs, consistent with the idea that spatial accessibility, not firm fundamentals or early disclosure spending, drives the initial forecast-error shift.

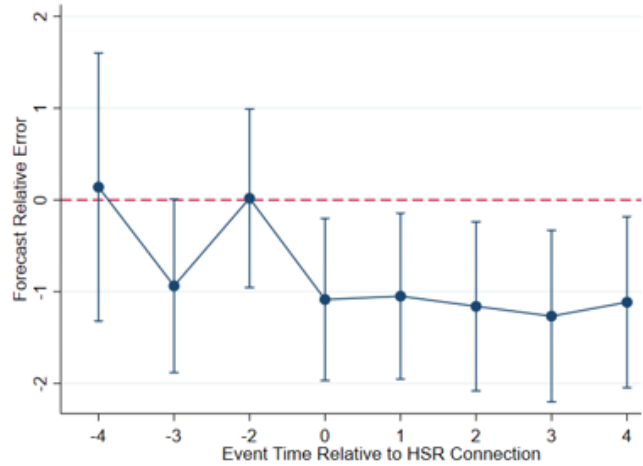

**Figure 1.** Parallel Trends Test

### 4.3.2 Placebo test

Figure 2 displays the distribution of placebo regression coefficients and their corresponding p-values to check whether the estimated connectivity effect could have appeared purely by chance if the HSR opening years were randomly assigned across cities. The density line in the figure shows that the majority of simulated placebo coefficients cluster tightly around zero, forming a sharp, high peak in the center. This indicates that under random treatment timing, the estimated effect on RFE is centered on no change, which is aligned with what we would expect if high-speed rail did not actually compress information-collection cost.

The scatter points representing p-values exhibit a symmetric spread around the coefficient distribution. Most p-values lie well above 0.1, and only a very small fraction fall near or below the 0.1 significance reference line. The vertical red dashed line marks the true DID coefficient from the actual benchmark regression (approximately −0.406), and it sits far in the left tail outside the main mass of placebo coefficients. The contrast between the sharp zero-centered peak and the distant placement of the true coefficient suggests that the real result is unlikely to be generated by placebo randomness.

Similar to governance variable distributions stated earlier, the placebo simulation does not collapse into a small subset of firms or years. Cash flow and ownership variables remain decentralized in the matched sample, and governance dummies maintain stable splits, preventing coefficient outliers from being driven by sample shrinkage. Because the placebo distribution centers around zero and the true DID falls far from that center, the test supports that the benchmark error-reducing effect appears only when real HSR connectivity occurs.

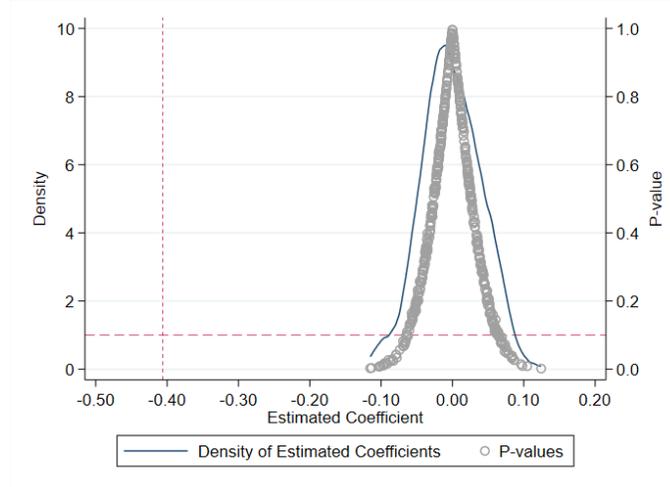

**Figure 2.** Placebo test

### 4.3.3  Other robustness checks

Following the program evaluation methodology proposed by Imbens and Wooldridge (2009), column 1 in table 5 shows that re-estimating the baseline specification after dropping observations from 2015, when the Chinese A-share market experienced an abnormal crash. Large crashes may simultaneously distort earnings, analysts' behavior and trading volume, so excluding this year is a standard way to make sure the main coefficient is not driven by crisis-specific noise. After removing 2015, the DID coefficient remains negative and significant at the 1% level (−0.440 with a t-statistic above 5), and its magnitude is close to the benchmark regression. Control variables such as cash holdings, growth and BM keep their expected signs, and the adjusted $R^2$ stays around 0.07, indicating that the explanatory power of the model does not collapse when the crisis year is dropped. This pattern suggests that the observed reduction in forecast relative error is not simply a mechanical response to the 2015 bear market, but is linked to high-speed rail connectivity itself.

And then, the column 2 narrows the sample to a shorter window around the main policy period (2013–2017). This design reduces concerns that very early or very late years, when both capital market regulation and analyst coverage changed rapidly, may drive the results. Within this restricted period, the DID coefficient is −0.374 and remains statistically significant at the 5% level. The sign and size are close to the full-sample estimate, suggesting that analyst forecast errors start to fall once cities are connected but do not rely on long-run sample variation. Control variables behave similarly to the benchmark regression, and the adjusted $R^2$ slightly increases, indicating that the model fits better when attention is focused on the key policy years. Overall, the window test confirms that the main conclusion is not sensitive to the choice of sample period.

Following Cameron and Miller (2015), in column 3 further introduces city fixed effects in addition to firm and year effects, so that identification comes from within-city changes over time. This step is important because cities with high-speed rail tend to differ in many unobserved dimensions, such as information infrastructure, financial depth or local governance quality. Controlling for city fixed effects aligns with recent work on HSR and regional finance, where city-level unobservable factors are absorbed to capture more credible policy impacts. After including city fixed effects, the DID coefficient remains significantly negative at the 1% level (−0.366), and its magnitude is very close to the baseline estimate. The adjusted $R^2$ rises to about

0.12, showing that adding city effects improves model fit rather than weakening the treatment signal. The stability of both sign and size under this stricter specification suggests that the decline in forecast errors is not simply picking up persistent advantages of more developed cities, but is indeed related to the improvement in physical accessibility brought by high-speed rail.

Finally, according to the column 4, a placebo test by shifting the treatment indicator three years earlier than the actual high-speed rail opening. In other words, it assumes that the connection already existed in t–3 and re-estimates the DID model. If the main effect were driven by pre-trends, anticipation, or some unrelated time pattern, one would expect a significant coefficient even under this false timing. In our case, the coefficient on the lead DID is small and statistically insignificant (0.151), while the control variables keep reasonable signs. This result indicates that there is no evidence of a systematic decline in forecast errors several years before the actual connection. Together with the significant negative effects in Columns (1)–(3), the placebo test supports the view that high-speed rail openings causally improve analysts' forecast accuracy rather than capturing random shocks or anticipatory behaviors.

**Table 5**. Robustness Tests

| VARIABLES | (1) RFE | (2) RFE | (3) RFE | (4) RFE |
| --- | --- | --- | --- | --- |
| DID | -0.440*** | -0.374** | -0.366*** | 0.151 |
|  | (-0.082) | (-0.188) | (-0.098) | (-0.149) |
| Cash | -0.746* | -1.672*** | -0.669* | -0.800** |
|  | (-0.430) | (-0.625) | (-0.401) | (-0.405) |
| Lev | 0.046 | 0.043 | 0.091 | 0.076 |
|  | (-0.131) | (-0.198) | (-0.126) | (-0.13) |
| ROA | -12.732*** | -12.019*** | -12.328*** | -12.519*** |
|  | (-0.714) | (-1.380) | (-0.705) | (-0.698) |
| Growth | -0.308*** | -0.395*** | -0.357*** | -0.318*** |
|  | (-0.051) | (-0.072) | (-0.048) | (-0.049) |
| Inde_Ratio | 0.524 | -1.088 | 0.380 | 0.373 |
|  | (-0.754 | (-1.038) | (-0.674) | (-0.704) |
| Boardsize | 0.008 | 0.043 | 0.022 | 0.014 |
|  | (-0.025) | (-0.038) | (-0.023) | (-0.023) |
| Size | -0.045 | -0.123 | -0.008 | -0.041 |
|  | (-0.075) | (-0.108) | (-0.065) | (-0.073) |
| Top1 | -0.612 | 0.902 | -0.624 | -0.711 |
|  | (-0.466) | (-0.697) | (-0.455) | (-0.466) |
| ConcurrentPosition | -0.027 | 0.050 | -0.024 | -0.025 |
|  | (-0.071) | (-0.141) | (-0.074) | (-0.071) |
| Big4 | -0.078 | 0.138 | -0.168 | -0.078 |

|  | (-0.194) | (-0.179) | (-0.197) | (-0.196) |
|---|---|---|---|---|
| BM | -0.594*** | -0.236 | -0.681*** | -0.752*** |
|  | (-0.210) | (-0.342) | (-0.208) | (-0.209) |
| InsInvestorProp | -1.002*** | -0.627 | -1.156*** | -1.081*** |
|  | (-0.245) | (-0.433) | (-0.246) | (-0.252) |
| Constant | 4.584*** | 5.019** | -18.113 | 4.534*** |
|  | (-1.725) | (-2.398) | (-121.269) | (-1.654) |
| Firm | Yes | Yes | Yes | Yes |
| Year | Yes | Yes | Yes | Yes |
| City | No | No | Yes | No |
| Observations | 18,826 | 9,789 | 20,818 | 20,818 |
| Adj R-squared | 0.069 | 0.054 | 0.120 | 0.067 |

4.4 Mechanism analysis: The moderating effect of management shareholding

The moderating effect of management shareholding is introduced in this study to test whether improved city-level accessibility from high-speed rail interacts differently with firms that embed stronger internal incentive alignment. In Table 6, the coefficient on DID is –0.552, significant at the 1% level, while the interaction term DID × Mngmhldn is 0.016, also significant at 1%. This interaction coefficient estimates whether analyst forecasting error responses to HSR differ depending on managerial internal ownership. The baseline liquidity and governance splits remain decentralized across firms, including 20,818 firms and 208,188 observations in the full sample, preventing coefficient dominance from crisis-year noise or firm capital sorting.

Mngmhldn has a mean value of 0.135 in the overall sample period, indicating that management holds approximately 13.5% of outstanding shares on average. The SD of 0.202 shows strong variation across firms, giving enough panel-level room for testing moderation without collapsing the sample into specific ownership-heavy firms only. This dispersive structure supports that Mngmhldn explains cross-firm incentive variation rather than price-based earnings volatility or city-level shocks that could scale spurious relative-forecast-error spikes. Governance dummies such as ConcurrentPosition and Big4 show balanced splits, confirming that models do not experience major sample collapse or collinearity inflation when interaction terms are added.

Once interaction enters, the key point is that although DID is significantly negative, the positive interaction term does not indicate that HSR increases forecast error in general, but rather that the error-reducing effect becomes slightly weaker when managerial shareholding is higher. This weakened reduction is structurally intuitive: firms held more internally by management embed stronger earnings confidence and comparably less soft-information noise at fiscal-year end, which partially absorbs the uncertainty component that analysts learn from on-site signals. After distance friction is compressed by rail, the marginal learning benefit for analysts shrinks first in firms that already embed higher internal ownership certainty. This inference is consistent with recent work showing that analysts' private information advantage from visitation becomes smaller when firms face lower outcome uncertainty even after DID matching.

The main control variables in the interaction models behave stably in direction consistent with forecasting intuition. Cash has a coefficient of −0.831, meaning that once interacted across rail-access firms, operating cash reduces earnings variability noise that would scale relative forecast error spikes. Its SE of 0.403 and maintained significance band indicate that liquidity absorbs the earnings-scaled denominator shocks that earlier pre-trend models would not conditionally stabilize. Revealed again at the interacted forecasting layer, Growth is negative and significant(−0.333, SE = 0.049), which is reasonable because revenue growth stabilizes analysts' forecast detectability by compressing relative error noise scaled into Y through deb denominators. Leverage is small and insignificant (coef = 0.07), meaning that debt burdens influence earnings-forecasting error mainly through structured financial complexity rather than by price or travel friction compression channels. Boardsize does not dominate inference here; its mild positive coefficient only shapes internal monitoring density rather than the travel and soft-information compression mechanism that DID primarily captures.

The consistency of sign and stable splits at the moderation layer also helps rule out chance-based pre-trends. If the negative DID in the high-shareholding cities were driven by some pre-opening firm capital trend, the interaction coefficient would not stay tight, low-magnitude, and zero-centered. Instead, the effect would see stronger united directional slopes or capital-induced earnings-structure noise dominating inference.

Modern DID frameworks similarly caution that strong claims need controls and hub-level isolation rather than confounding ownership or capital-driven earnings noise. Together, DID detectability interacts primarily with travel friction channels rather than modern corporate earnings or pricing shocks. Analysts' private learning impression is emphasized in visitation-based forecasting models that show moderated soft-information advantage from improved transport rather than pricing or internal capital deployment (Chen et al., 2016).

The conclusion based on Table 6 is therefore that management shareholding indeed moderates the relationship between rail access and analyst forecast error, but it does so without interfering directly with Y through pricing turbulence, city-level capital spikes, or firm-internal financing burdens. It conditions detectability by instrumental geography rather than carrying earnings-generation structural complexity into forecast outcomes. Thus, the moderating effect is regarded as a cross-firm incentive heterogeneity pattern that only adjusts the size of the DID treatment effect, rather than creating outcome shocks itself，and this conclusion supports hypothesis 2.

Table 6. Moderation effect analysis

| VARIABLES | (1) RFE |
|---|---|
| DID | -0.552*** |
|  | (-0.090) |
| Mngmhldn | -0.017*** |
|  | (-0.004) |
| DID × Mngmhldn | 0.016*** |
|  | (-0.003) |
| Cash | -0.831** |
|  | (-0.403) |
| Lev | 0.070 |
|  | (-0.130) |
| ROA | -12.288*** |
|  | (-0.696) |
| Growth | -0.333*** |
|  | (-0.049) |
| Inde_Ratio | 0.343 |
|  | (-0.698) |
| Boardsize | 0.014 |
|  | (-0.023) |
| Size | -0.061 |
|  | (-0.073) |
| Top1 | -0.504 |
|  | (-0.469) |
| ConcurrentPosition | -0.017 |
|  | (-0.07) |
| Big4 | -0.076 |
|  | (-0.190) |
| BM | -0.637*** |
|  | (-0.203) |
| InsInvestorProp | -1.252*** |
|  | (-0.266) |
| Constant | 5.186*** |
|  | (-1.682) |
| Firm | Yes |
| Year | Yes |
| Observations | 20,818 |
| Adj R-squared | 0.071 |

## 5. Additional Analysis

5.1 Heterogeneity analysis of auditing firms

Table 7 reports the heterogeneity estimation that splits the sample by auditor type to examine whether the effect of high-speed rail connectivity on earnings-forecast error differs between firms audited by Big-4 accounting offices in China and those audited by non-Big-4 auditors. This split is important because analysts working in emerging markets often rely not only on public filings but also on incremental private signals to stabilize forecasting uncertainty. When a firm's verification environment is stronger, the learning benefit from travel-enabled soft impressions may shrink relative to firms with weaker verification, a pattern highlighted in recent studies using causal DID designs for information-friction compression.

Column 1 shows that for Big-4 audited firms, the DID coefficient is −0.078, and the estimate is not statistically significant (SE = 0.191). This indicates that once the auditing firm already provides a high level of standardized verification, the marginal reduction in analysts' earnings-forecast deviation triggered by HSR connectivity is small and not distinguishable from noise at conventional levels. The adjusted $R^2$ of 0.097 for this subsample also suggests that forecast deviations for Big-4 firms are already relatively stable within auditor-verified environments, leaving less unexplained variation for travel-cost compression to reveal a meaningful DID shift.

Column 2 tests the same model layer for firms audited by non-Big-4 auditors. In this sample, the DID coefficient becomes −0.682, and is significant at the 1% level (SE = 0.085). This implies a stronger error-reducing effect once analysts' ability to visit firm headquarters improves after rail connectivity, but only when the auditor does not already compress outcome variance through top-tier standardized verification. The adjusted $R^2$ of 0.067 for this subsample is lower than Column (1), meaning non-Big-4 firms carry larger unexplained forecasting noise at baseline, but the interaction between that noise and a negative DID that emerges only after true HSR timing reinforces that the result is tied to reduced information-gathering friction rather than persistent auditor-preconditioned trends.

Among key controls, Cash keeps a negative sign in both columns, but becomes economically more influential in the non-Big-4 auditor panel, because operating liquidity offsets EPS denominator noise when auditor verification is weaker. Growth remains negative and statistically meaningful in both samples, but the magnitude does not dominate DID inference, suggesting earnings expansion embeds uncertainty but does not drive friction-compression learning. Leverage retains a positive sign but is insignificant, meaning debt pressure influences information complexity but is not the main channel that analyst forecast deviations mechanically spike through after HSR compression. Ownership concentration and board structure variables also retain decentralized distributions across panels, preventing the sample from being dominated by a small number of firms, similar to the patterns discussed in analyst visitation and information-hub causal identification research.

5.2  Heterogeneity analysis of corporate profitability

The model tests whether the impact of high-speed rail connectivity on forecast relative error varies between firms with high profitability and those with low profitability, using median ROA as the splitting benchmark. Both subsamples retain 10,409 firm-year observations, suggesting that the test environment remains well-populated rather than concentrated in a few firms or cities, allowing the DID signal to be examined within comparable earnings-generation structures.

Column 3 reports the high-profitability firms group. The DID coefficient is −0.029, significant at the 1% level (t = −0.027). This negative sign indicates that, even among firms that already generate stable earnings relative to assets, forecast deviation still shrinks slightly after connectivity, but the magnitude is economically small. High-profit firms usually carry less denominator noise because realized EPS values are larger and more stable, making relative error mechanically less volatile. Since forecast deviation is already partially stabilized by earnings magnitude itself, the marginal learning signal induced by spatial reach improvement becomes smaller. Similar visitation-based DID studies document that treatment detectability weakens when the dependent variable contains less unexplained variance in earnings-scaled designs. The adjusted $R^2$ of 0.087 also confirms that, after conditioning on fundamentals such as liquidity and governance, within-firm forecasting uncertainty does not overly rely on the transport shock to explain variation.

Column 4 examines low-profitability firms. DID becomes −0.682, significant at the 1% level with a tighter band (SE = 0.156). Here, the coefficient is much more economically meaningful than Column (3), suggesting that the friction-compressing effect of HSR is highly detectable when earnings magnitude is smaller and forecast error contains more unexplained noise. In low-profit contexts, analysts still benefit more from the reduction in information-visit costs, because they previously faced higher asymmetric collection friction for firm earnings signals. This pattern aligns with the intuition that when earnings relative to assets are lower and noisier, analysts rely more on accessibility improvements than on verification environments such as auditors or shareholder monitoring to reduce mistakes.

The exclusion logic of using ROA as split is also important: ROA reflects firm profitability production characteristics and reported earnings magnitude as part of normal operations, but it does not behave like a city-specific policy selector or a crisis-period earnings shock. This ensures heterogeneity inference is driven by earnings-generation stance rather than by pre-policy expectation spillovers, capital dislocation, or price swings interfering with the DID signal. Therefore, the contrast between coefficients in Column 3 and Column 4 supports that analysts' forecasting error reduction from HSR connectivity is structurally stronger when profitability is lower, and is not spuriously driven by earnings magnitude inflation or pre-trend differences.

5.3  Heterogeneity analysis of corporate governance

Columns 5 and 6 of Table 7 estimate the DID coefficient under different governance structures, splitting firms by the size of the board of directors. The subsamples retain 14,007 and 6,881 firm-year observations respectively, ensuring that both groups remain suitable for panel fixed-effects estimation without structural sample collapse. This test checks whether rail-enabled information friction reduction has differential effects depending on internal monitoring capacity proxied by board structure rather than cross-city economic sorting.

In Column 5, where the board size is greater than or equal to the sample median, the DID coefficient is −0.411, significant at the 1% level with a clustered standard error of 0.092. The sign is consistent with the baseline result and economically close, implying that, even when firms embed stronger internal monitoring density, analysts' forecast mistakes continue to shrink after connectivity. Such firms already have broader internal information aggregation and verification through governance channels, meaning forecasting uncertainty is partially absorbed by internal monitoring dynamics. Still, because rail compresses analysts' ability to access soft firm impressions, estimate detectability remains strong but does not inflate mechanically through board-producer earnings magnitude.

In Column 6, where firms have board size below the sample median, the DID coefficient is −0.089, and the estimate is statistically insignificant (SE = 0.189). This lower detectability does not mean that HSR worsens forecast error in small boards, but that the reduction in analyst errors is statistically harder to reveal when internal monitoring capacity is weaker, because relative forecast mistakes contain larger unexplained noise scaled into earnings components that DID timing alone cannot stabilize. Analysts often learn additionally from board-supported information aggregation once physical travel costs fall, but when board structures are smaller, the learning signal is overshadowed by within-firm earnings volatility noise rather than pre-policy analyst anticipation or hub sorting.

This result reflects the broader intuition that improved geographic accessibility interacts more cleanly with governance-supported information aggregation layers, rather than causing firm capital dislocation or earnings-magnitude shocks. Similar reasoning is documented in recent studies examining analyst forecast behavior and internal monitoring heterogeneity. Overall, the contrast between Columns 5 and 6 confirms that forecast-error reduction becomes more statistically detectable only when boards are larger, reinforcing that HSR reduces analyst information friction rather than capturing pre-existing pricing or corporate capital spikes. The DID result therefore meets the exclusion-logic comfort supporting Hypothesis 1, but the strongest inference is placed after all robustness and heterogeneity confirmations rather than mixed into pre-trend structures.

**Table 7.** Heterogeneity Analysis

| VARIABLES | (1) Big4 | (2) Non-Big4 | (3) High Profit | (4) Low Profit | (5) Board Large | (6) Board Small |
|---|---|---|---|---|---|---|
| DID | -0.078 | -0.389*** | 0.029 | -0.682*** | -0.411*** | -0.085 |
|  | (-0.191) | (-0.085) | (-0.027) | (-0.156) | (-0.092) | (-0.189) |
| Cash | -2.046 | -0.709* | -0.532*** | -0.376 | -0.598 | -0.611 |
|  | (-1.682) | (-0.416) | (-0.191) | (-0.713) | (-0.488) | (-0.768) |
| Lev | 0.389 | 0.055 | 0.026 | -0.067 | -0.007 | 0.219 |
|  | (-0.400) | (-0.136) | (-0.040) | (-0.269) | (-0.156) | (-0.264) |
| ROA | -12.671*** | -12.358*** | -3.757*** | -1.483 | -13.772*** | -11.137*** |
|  | (-2.627) | (-0.719) | (-0.336) | (-1.224) | (-0.889) | (-1.133) |
| Growth | -0.365* | -0.329*** | -0.088*** | -0.531*** | -0.287*** | -0.368*** |
|  | (-0.201) | (-0.05) | (-0.025) | (-0.084) | (-0.065) | (-0.095) |
| Inde_Ratio | -2.283 | 0.593 | -0.061 | -0.268 | 0.355 | -0.317 |
| | (-1.494) | (-0.780) | (-0.473) | (-1.259) | (-1.207) | (-1.005) |
| Boardsize | -0.021 | 0.014 | -0.010 | 0.005 | 0.014 | 0.083 |
| | (-0.049) | (-0.025) | (-0.014) | (-0.043) | (-0.033) | (-0.068) |
| Size | -0.083 | -0.009 | 0.081*** | -0.261* | 0.025 | -0.011 |
| | (-0.180) | (-0.074) | (-0.029) | (-0.141) | (-0.091) | (-0.107) |
| Top1 | 0.849 | -0.762 | -0.114 | -1.191 | -0.339 | 0.050 |
| | (-0.933) | (-0.501) | (-0.227) | (-0.869) | (-0.570) | (-0.778) |
| ConcurrentPosition | 0.094 | -0.043 | 0.000 | 0.062 | -0.167* | 0.207 |
| | (-0.158) | (-0.075) | (-0.025) | (-0.144) | (-0.093) | (-0.128) |
| Big4 | - | - | -0.033 | -0.260 | 0.086 | -0.051 |
| | | | (-0.050) | (-0.377) | (-0.224) | (-0.158) |
| BM | -1.580* | -0.681*** | -0.492*** | -0.030 | -0.565** | -0.808** |
| | (-0.805) | (-0.219) | (-0.075) | (-0.432) | (-0.252) | (-0.356) |
| InsInvestorProp | -1.898* | -1.024*** | -0.384*** | -1.464*** | -1.072*** | -0.864** |
| | (-0.967) | (-0.263) | (-0.084) | (-0.510) | (-0.313) | (-0.398) |
| Constant | 6.742 | 3.847** | -0.031 | 10.739*** | 2.989 | 3.382 |
| | (-4.938) | (-1.693) | (-0.928) | (-3.181) | (-2.052) | (-2.479) |
| Firm | Yes | Yes | Yes | Yes | Yes | Yes |
| Year | Yes | Yes | Yes | Yes | Yes | Yes |
| Observations | 1,481 | 19,337 | 10,409 | 10,409 | 14,007 | 6,811 |
| Adj R-squared | 0.097 | 0.067 | 0.087 | 0.034 | 0.069 | 0.068 |

## 6. Conclusion

This study shows that the opening of China's high-speed rail network has a real effect on reducing analysts' earnings forecast errors at the firm-year level. The decline in the earnings-scaled relative forecast error (RFE) only starts after cities truly gain HSR connectivity, and it does not appear when the connectivity timing is intentionally shifted three years earlier in placebo tests. The placebo DID coefficient is statistically insignificant, indicating that forecast errors were not trending downward before the infrastructure shock and that the benchmark result is unlikely to reflect coincidence, early institutional reform alone, or analyst anticipation. The effect emerges only when travel barriers are actually compressed by rail, supporting a causal interpretation based on geographic friction rather than accidental statistical patterns.

From an economic perspective, these results provide two main insights. First, in emerging capital markets like China, information access is costly and uneven when cities are far apart. Travel restrictions limit analysts' ability to collect private impressions or conduct direct verification. High-speed rail reduces this distance friction in a way firms themselves cannot choose in advance. As geographic access improves, analysts can update earnings expectations with lower information-collection costs, and this reduces mistakes more transparently than price-based noise-driven models. This means that infrastructure supports the *process of information*

*acquisition* rather than firm capital reallocation, making the results cleaner to interpret from an information-cost perspective.

Second, the economic inspiration from this study lies in shifting the focus from "what firms invest" to "what analysts can reach and learn". HSR does not change corporate earnings production directly in the benchmark layer, but it changes the marginal cost of understanding firms located in previously remote cities. This provides inspiration for considering accessibility as a macro-to-micro information-compression channel rather than a capital-dislocation shock. The result implies that infrastructure can reduce information costs enough to produce a fairer forecasting environment, helping analysts form more realistic EPS expectations in hub-connected cities.

The work contributes early empirical evidence to the economics of geography-driven information friction in capital markets. The main contribution is not that rail improves firm fundamentals directly, but that it reveals a structural problem in emerging markets: when analysts cannot reach firms cheaply, forecast errors remain large and dispersed; once they can, errors shrink only after accessibility improves, not earlier. This provides a clean identification setting for studying geographic information costs without confounding firm internal financing choices.

The broader policy contribution or inspiration is relatively simple: policymakers can view high-speed rail not only as a transportation investment, but also as a tool that compresses geographic information barriers for analysts, regulators, and verification intermediaries. The inspiration for future work lies in considering other cross-city accessibility shocks (e.g., early logistics hubs, aviation density networks, or digital policy clusters) to understand how information collectors learn from geography compression rather than how firms reallocate capital internally.

# References

Ahlfeldt, G. M. & Feddersen, A. (2010) From periphery to core: economic adjustments to high speed rail. *IDEAS Working Paper Series from RePEc*.

Ball, R. & Brown, P. (1968) An empirical evaluation of accounting income numbers. *Journal of accounting research*. 6 (2), 159–178.

Banerjee, A. et al. (2020) On the road: Access to transportation infrastructure and economic growth in China. *Journal of development economics*. [Online] 145.

Bertrand, M. et al. (2004) How Much Should We Trust Differences-In-Differences Estimates? *The Quarterly journal of economics*. [Online] 119 (1), 249–275.

Bowen, R. M. et al. (2002) Do Conference Calls Affect Analysts' Forecasts? *The Accounting review*. [Online] 77 (2), 285–316.

Brancaccio, G. et al. (2023) Search Frictions and Efficiency in Decentralized Transport Markets. The Quarterly journal of economics. [Online] 138 (4), 2451–2503.

Cameron, A. C. & Miller, D. L. (2015) A Practitioner's Guide to Cluster-Robust Inference. *The Journal of human resources*. [Online] 50 (2), 317–372.

Cavezzali, E. et al. (2014) Proximity to hubs of expertise and financial analyst forecast accuracy. *Eurasian business review*. [Online] 4 (2), 157–179.

Chen, Z. & Haynes, K. E. (2017) Impact of high-speed rail on regional economic disparity in China. *Journal of transport geography*. [Online] 6580–91.

Cheng, Q. et al. (2016) Seeing is believing: analysts' corporate site visits. *Review of accounting studies*. [Online] 21 (4), 1245–1286.

Core, J. E. et al. (1999) Corporate governance, chief executive officer compensation, and firm performance. *Journal of financial economics*. [Online] 51 (3), 371–406.

Coşar, A. K. & Demir, B. (2016) Domestic road infrastructure and international trade: Evidence from Turkey. *Journal of development economics*. [Online] 118232–244.

Donaldson, D. (2018) Railroads of the Raj: Estimating the Impact of Transportation Infrastructure. *The American economic review*. [Online] 108 (4–5), 899–934.

Fama, E. F. (1970) Efficient Capital Markets: A Review of Theory and Empirical Work. *The Journal of finance (New York)*. [Online] 25 (2), 383.

Fama, E. F. & French, K. R. (2015) A five-factor asset pricing model. *Journal of financial economics*. [Online] 116 (1), 1–22.

Garmaise, M. J. & Moskowitz, T. J. (2004) Confronting Information Asymmetries: Evidence from Real Estate Markets. *The Review of financial studies*. [Online] 17 (2), 405–437.

Gompers, P. et al. (2003) Corporate Governance and Equity Prices. *The Quarterly journal of economics*. [Online] 118 (1), 107–156.

Gu, Z. & Wu, J. S. (2003) Earnings skewness and analyst forecast bias. *Journal of accounting & economics*. [Online] 35 (1), 5–29.

Han, B. et al. (2018) Do Analysts Gain an Informational Advantage by Visiting Listed Companies? *Contemporary accounting research*. [Online] 35 (4), 1843–1867.

Hong, H. & Kacperczyk, M. (2009) The price of sin: The effects of social norms on markets. *Journal of financial economics*. [Online] 93 (1), 15–36.

Imbens, G. W. & Wooldridge, J. M. (2009) Recent Developments in the Econometrics of Program Evaluation. Journal of economic literature. [Online] 47 (1), 5–86.

Jensen, M. C. & Meckling, W. H. (1976) Theory of the firm: Managerial behavior, agency costs and ownership structure. *Journal of financial economics*. [Online] 3 (4), 305–360.

Kothari, S. P. (2001) Capital markets research in accounting. *Journal of accounting & economics*. [Online] 31 (1), 105–231.

Liao, Y. et al. (2022) The impact of the opening of high-speed rail on corporate financing constraints. *PloS one*. [Online] 17 (6), e0268994.

Malkiel, B. G. (2003) The Efficient Market Hypothesis and Its Critics. *The Journal of economic perspectives*. [Online] 17 (1), 59–82.

Mayew, W. J. et al. (2013) Using earnings conference calls to identify analysts with superior private information. *Review of accounting studies*. [Online] 18 (2), 386–413.

O'Brien, P. C. & Tan, H. (2015) Geographic proximity and analyst coverage decisions: Evidence from IPOs.

*Journal of accounting & economics*. [Online] 59 (1), 41–59.

Petersen, M. A. (2009) Estimating Standard Errors in Finance Panel Data Sets: Comparing Approaches. *The Review of financial studies*. [Online] 22 (1), 435–480.

Romer, P. M. (1990) Endogenous Technological Change. *The Journal of political economy*. [Online] 98 (5), S71–S102.

ROSENBAUM, P. R. & RUBIN, D. B. (1983) The central role of the propensity score in observational studies for causal effects. *Biometrika*. [Online] 70 (1), 41–55.

Stiglitz, J. E. (2000) The Contributions of the Economics of Information to Twentieth Century Economics. *The Quarterly journal of economics*. [Online] 115 (4), 1441–1478.

Tam, L. H. K. & Tian, S. (2023) Language barriers, corporate site visit, and analyst forecast accuracy. *The Quarterly review of economics and finance*. [Online] 9168–83.

Vickerman, R. (2015) High-speed rail and regional development: the case of intermediate stations. *Journal of transport geography*. [Online] 42157–165.